# A Fast Audio Clustering Using Vector Quantization and Second Order Statistics

*Konstantin Biatov*

konst.biatov@googlemail.com

## Abstract

This paper describes an effective unsupervised speaker indexing approach. We suggest a two stage algorithm to speed-up the state-of-the-art algorithm based on the Bayesian Information Criterion (BIC). In the first stage of the merging process a computationally cheap method based on the vector quantization (VQ) is used. Then in the second stage a more computational expensive technique based on the BIC is applied. In the speaker indexing task a turning parameter or a threshold is used. We suggest an on-line procedure to define the value of a turning parameter without using development data. The results are evaluated using ESTER corpus.

**Index Terms**: speaker indexing, speaker clustering, Bayesian Information Criterion, vector quantization

## 1. Introduction

The unsupervised speaker indexing task is to group together audio segments or utterances belonging to the same speaker in the large audio collection [1]. The application of speaker indexing in the audio database consists of organizing the audio data according to the speakers presented in the database. A fast speaker indexing is important for real-time application in particular for speaker retrieval and adaptation. The unsupervised speaker indexing is closely related with the unsupervised speaker clustering techniques.

The speaker clustering task has been in the focus of researchers in the last decade. Most of the state-of-the-art systems in the speaker clustering use second-order statistics such as the BIC [2], the Kullback-Leibler distance (KL2) [3], the Hotelling $T^2$ statistic [4], the generalized likelihood ratio (GLR) [5], the cross likelihood ratio (CLR) [6] and also the Information Bottleneck (IB) principle [7] the variational Bayesian methods [8], the speaker factors analysis [9] and some other techniques. In [10] a flexible framework that selects an optimal speaker model, Gaussian Mixture Models (GMM) or VQ, based on BIC according to the duration of the utterances is described. Currently some interest in the fast speaker clustering strategy has been indicated in the recent publications [7], [11]. The proposed in [7] system is based on the IB principle and can achieve the diarization error rate of 23.2% using for evaluation meeting data. The error rate is comparable with the Hidden Markov Models (HMM)/GMM baseline system but running faster: 0.3xRT. In [11] the basic idea is to adopt a computationally cheap method to reduce the hypothesis space of the more expensive and accurate model selection via the BIC. Two strategies based on the pitch-correlogram and the unscented-transform based approximation of the KL divergence are used independently as a fast-match approach to select the most likely clustering merge. The new system speeds up the diarization by 41% and is running 0.88xRT without affecting the diarization error rate. The fast clustering technique becomes practical for the speaker indexing in the large audio database.

Many approaches applied to speaker clustering use a threshold for merging segments in one cluster. Usually this threshold is evaluated using development data. The value of the threshold, for example for the BIC based clustering, varied from one application to another significantly depending on the used features and their dimensionality. For example, in [5] 12 dimensional Mel cepstral Frequency coefficient (MFCC) feature vectors are used and the value of the threshold $\lambda$ is 12.0. In [12] 26 MFCCs (12 MFCCs + energy + their first derivatives) are used and $\lambda$ was defined as 3.0. The computation of an adaptive threshold $\lambda$ for unsupervised speaker segmentation using the BIC is presented in [13]. The turning parameter is corrected depending on the size of the window used in the BIC. Usually for threshold definition the development data are required. The threshold definition is a well known task, in particular in speaker verification applications. Some techniques of threshold definition in speaker verification are presented in [14], [15]. Another approach of computationally efficient strategy for setting a priori threshold in the adaptive speaker verification system is presented in [16].

The speaker clustering task usually includes two tasks: audio segmentation into speaker homogeneous segments and merging segments belonging to the same speaker in one cluster. In this paper we will focus on fast merging criterion to speed-up the clustering process. In each of the iteration of the agglomerative clustering process we exploit a two steps procedure. In the first fast step N-closest candidates segments for merging are obtained. Then in the second step these N segments are compared and then the two closest segments are selected for merging using a more computationally expensive procedure such as the BIC. In the second step the merging criterion is not limited by using the BIC. The other distance measure based on the CLR, the GLR or other distance measures could be applied. The suggested method could be also applied to the case when segments are modeled by the GMM or adapted GMM instead of one Gaussian model with full covariance as it is used in the BIC. The first step provides fast match and the second step is used for final precise selection of the merging segments. The described strategy is applied to an agglomerative speaker clustering (ASC) task. The suggested technique can reduce the speaker clustering time comparing with the state-of-the-art clustering and provides speaker clustering much faster than real-time.

In speaker clustering, in particular, in the BIC based clustering the merging criterion depends on the threshold $\lambda$ that is heuristically determined depending on the used features. This threshold can be obtained manually or using some validation data to find optimal value for a given data set. In this paper we suggest on-line procedure using actual clustering data.

The rest of the article is organized as follows: section 2 describes a state-of-the-art algorithm for speaker clustering, section 3 explains the suggested algorithm for speaker clustering, section 4 explains the on-line algorithm to obtain the threshold, section 5 presents the data sets used in the experiments, section 6 describes evaluation metrics, section 7 describes experiments and section 8 summarizes this article and points out the future work.

## 2. The BIC base speaker clustering

The BIC for audio application was initially proposed in [2]. In general the BIC is defined as:

$$BIC(M) = \log L(X,M) - 0.5 \lambda \#(M)\log(N), \quad (1)$$

where $\log L(X,M)$ denotes likelihood of segment X given by the model M, N is the number of feature vector in the data, $\#(M)$ is the number of free parameter in the model and $\lambda$ is a tuning parameter. For Gaussian distributions in order to estimate the data distribution turn point between two segments $i$ and $j$ that have $n_i$ and $n_j$ frames respectively, the $\Delta$BIC value is computed as:

$$\Delta BIC = 0.5 n_i \log|\Sigma_i| + 0.5 n_j |\Sigma_j| -$$
$$0.5 n_{ij} \log|\Sigma_{ij}| + \lambda P \quad (2)$$

where $n_{ij} = n_i + n_j$, d is the dimension of the feature vector, $\Sigma_{ij}$ is the covariance matrix of the data points from two segments $i$ and $j$, $\Sigma_i$ is the covariance matrix of the data points from the segment $i$, $\Sigma_j$ is the covariance matrix of the data points from segment j and P is the penalty:

$$P = 0.5(d+0.5d(d+1))\log(n_i + n_j). \quad (3)$$

The $\Delta$BIC is the distance between two Gaussian models. The negative value of the $\Delta$BIC indicates that two models fit to the data better than one common Gaussian model. The positive value of the $\Delta$BIC indicates statistical similarity of the compared models. In the ASC, the $\Delta$BIC is used to make a pair-wise comparison between audio segments. If the $\Delta$BIC corresponding to the two segments is positive and maximal with respect to the other pairs of segments these two segments are merged. The comparison continues until there are no more pairs of the segments with the positive $\Delta$BIC.

## 3. Speaker Clustering Based on VQ and BIC

In this paper we suggest an effective algorithm that uses the VQ and the BIC for speaker clustering. In the ASC, the BIC should be applied to test all pairs of segments to select the two best for merging. The complexity of this algorithm in each iteration with respect to the BIC calculation is $O(n^2)$ where n is the actual number of segments. With the relatively large number of the speech segments the speaker clustering procedure based on the BIC requires heavy computational cost. The VQ technique previously was successfully used for speaker recognition task [17]. The VQ for comparison of short audio utterance was suggested in [10].

In the preprocessing phase the MFCCs are extracted. Before the ASC, the GMM is estimated on the entire audio data that should be clustered. The number of the GMM corresponds to the number of speech segments. The means of the GMMs are exploited as the cluster's centroids and are used for the VQ. Each feature vector of speech segment is compared with the cluster centroids that are the means of the GMMs. Then the nearest centroid is calculated using Euclidean distance. The number of the nearest cluster centroid is used as a codebook number. For each speech segment the histogram of the codebook number is calculated. For each audio segment the frequencies of the cluster numbers are normalized by the sum of these frequencies. The obtained normalized histograms are considered as secondary feature vectors of the audio segments. The distance between audio segments are calculated using cosine distance between normalized histograms that correspond to the compared segments. Each audio segment is described as by the frame-based MFCC vector and by the normalized histogram. We suggest a two step merging procedure in the ASC algorithm. The suggested speaker clustering algorithm is performed as follows:

1. Every initial segment in the beginning of the ASC is represented by normal probability distribution function with mean vector and full covariance matrix and by the normalized histogram, obtained using vector quantization procedure.
2. The merging procedure includes two steps. In the first step all pairs of segments are compared based on histogram vectors using cosine distance. The N-closest pairs are selected. In the second step, $\Delta$BIC is calculated for all N-best selected pairs to find one closest for merging.
3. Then the models and the histogram vector corresponding to the merging segments are updated.
4. The process is repeated until no more segments pairs remain to be merged. The stop criterion is based on the $\Delta$BIC.

The computational complexity of this process is compounded from two parts. From the computational complexity of the histogram vector comparison, using cosine distance and the computational complexity of the BIC based comparison of the small number of the N most nearest segments. The number N $<< n^2$ where n is the actual number of the segments. The computational complexity of the histogram vectors comparison using cosine distance is quite low. The use of the $\Delta$BIC for a limited number of segments is also computationally not expensive. Instead of the BIC the GLR, the CLR, the KL2, the arithmetic harmonic sphericity measure (AHS), the divergence shape (DS), the Hotelling $T^2$ statistic, the Bhatacharyya distance or other metrics could be used.

## 4. On-line Threshold Calculation

In the speaker clustering task, in particular, in the BIC based clustering the merging criterion depends on the threshold $\lambda$ that is heuristically determined depending on the used features. Even BIC criterion uses penalty, the turning parameter should be specially adjusted depending on the data dimensionality to obtain better results. This threshold can be obtained using some validation data to find optimal value for a given data set. The optimal value of tuning parameter varies significantly depending on the data dimensionality. The on-line BIC threshold correction in the speaker segmentation task is suggested in [13]. The correction is based on the size of the window used in the BIC. In the unsupervised speaker clustering task it is not defined which speakers belong to the same cluster and which speakers are from the different clusters. In the speaker clustering task we suppose that each audio segment is homogeneous and belongs to one speaker. We suggest for the threshold evaluation to compare parts of the same segment. Each audio segment *i* available for the clustering is divided into two parts. For each segment *i* and for their two parts the *ΔBIC* is calculated. The *ΔBIC* should be positive since these two parts are from the same segment. Resolving inequality the value of $\lambda$ could be obtained as

$$\lambda > (0.5 n_i \log |\Sigma_i| - 0.5 n_{i1} \log |\Sigma_{i1}| - 0.5 n_{i2} |\Sigma_{i2}|) P^{-1} \quad (4)$$

where $n_i = n_{i1} + n_{i2}$, $d$ is the dimension of the feature vector, $\Sigma_i$ is the covariance matrix of the data points from segments $i$, $\Sigma_{i1}$ is the covariance of the data points of the left part of the segment i, $\Sigma_{i2}$ is the covariance matrix of the data points from the right part of the segment $i$ and $P$ is penalty. For each segment that should participate in the clustering process the value of the $\lambda$ is obtained. Then the average value of $\lambda$ namely $\ddot{\lambda}$ and their standard deviation $\sigma$ are calculated. In the speaker verification task some approaches are proposed to automatically estimate the speaker dependent threshold a priori [15]. It is suggested to estimate threshold using data only from clients, without data from imposters or other additional development data. The threshold is estimated as linear combination of clients score, standard deviation and some coefficients that should be obtained empirically. Exploiting this idea we suggest calculating the actual value for turning parameter $\lambda$ as

$$\lambda_{act} = \alpha \ddot{\lambda} + \beta \sigma \quad (5)$$

where $\ddot{\lambda}$ is average value of the $\lambda$, $\sigma$ is standard deviation of the $\lambda$ calculated using data enrolled for the clustering. The constants $\alpha$ and $\beta$ should be obtained empirically. They are defined once and are valid for all type of the data. Experimentally the value of $\alpha$ is defined as 2 and value of $\beta$ is defined as 0.5.

## 5. Database Description

For speaker clustering evaluation two databases are used. The first part of the evaluation data is from ELRA [18]. This collection is the corpus collected in the French national project ESTER (Evaluation of Broadcast News enriched transcription systems) [19]. The original transcriptions contain a total of 2,172 different speakers. About one third (744) are female speakers while 1,398 are male speakers. The data obtained from the ELRA (www.elra.info) includes 517 speakers and 47880 audio segments. For experiments 10 audio files from the ESTER collection with the total duration of 10 hours are used. The data is converted into 16kHz sampling rate and 16 bit per second. The audio signals are divided into overlapping frames with the size 25 ms. An overlap is 10 ms. The features consist of all-purposes MFCC. In the experiments we use 12 MFCC, energy and their first derivatives.

## 6. The Evaluation Metric

The performance of the speaker clustering is measured via mapping between reference labels and system hypothesis. We use purity-based evaluation as was described in [20]. The quality of the system is evaluated using both speaker purity (SP) and cluster purity (CP) based on the segments and on the frames level. On the segment level purity evaluation the size of the segments is not taken into account. It is done to equalize the input of large and small segments into purity value.

## 7. Experimental Results

For evaluation of the suggested approach some experiments are carried out. In the experiments we compare the results of clustering using the state-of-the-art algorithm and suggested algorithm. The suggested algorithm is evaluated using different N-closest segments in the first stage of the merging process. The value of N is 100, 200, 300, 400 and 500. The speaker purity, cluster purity based on segment and frame levels and clustering time for the ESTER corpus are presented on Table 1 and Table 2. Table 3 presents 10 files from the ESTER corpus including duration of the files, number of actual speakers in the file, number of segments in the files and number of the clusters obtained using suggested algorithm and obtained using BIC based clustering. The first number of cluster column shows the number of clusters obtained using suggested algorithm and the second number shows the number of clusters obtained using baseline algorithm.

Table 1. *Evaluation of clustering results of the ESTER corpus using segment level purity.*

| Clustering method | SP | CP | Time in xRT |
|---|---|---|---|
| BIC | 0.831 | 0.805 | 0.22 |
| VQ-BIC 100-best | 0.815 | 0.850 | 0.03 |
| VQ-BIC 200-best | 0.843 | 0.840 | 0.04 |
| VQ-BIC 300-best | 0.849 | 0.839 | 0.05 |
| VQ-BIC 400-best | 0.843 | 0.830 | 0.05 |
| VQ-BIC 500-best | 0.846 | 0.824 | 0.06 |

Table 2. *Evaluation of clustering results of the ESTER corpus using frame level purity.*

| Clustering method | SP | CP | Time in xRT |
|---|---|---|---|
| BIC | 0.765 | 0.787 | 0.22 |
| VQ-BIC 100-best | 0.741 | 0.771 | 0.03 |
| VQ-BIC 200-best | 0.767 | 0.794 | 0.04 |
| VQ-BIC 300-best | 0.768 | 0.795 | 0.05 |
| VQ-BIC 400-best | 0.768 | 0.794 | 0.05 |
| VQ-BIC 500-best | 0.768 | 0.795 | 0.06 |

Table 3. *Clustering results of the ESTER corpus*

| Files | Speakers | Clusters | Segments | Duration (sec.) |
|---|---|---|---|---|
| 1 | 43 | 32/32 | 109 | 3666 |
| 2 | 24 | 24/24 | 102 | 3549 |
| 3 | 47 | 41/41 | 118 | 3603 |
| 4 | 39 | 34/34 | 119 | 3595 |
| 5 | 43 | 34/34 | 100 | 3625 |
| 6 | 29 | 27/26 | 112 | 3629 |
| 7 | 45 | 39/39 | 122 | 3618 |
| 8 | 42 | 42/42 | 131 | 3625 |
| 9 | 20 | 23/23 | 103 | 3654 |
| 10 | 43 | 40/40 | 117 | 3603 |

Table 4 presents the experiments of the on-line BIC threshold tuning. The threshold is data dependent and is evaluated on-line using data enrolled for the clustering. For clustering BIC is used. In the experiments four feature sets with different dimensionality are tested using 10 hours audio data. Table 4 presents average speaker/cluster purity on the segment level, on the frame level and average value of actual threshold. The experiments are carried out on a computer with processor Intel (R), Core™ 2 Q8300, 2.5 GHz.

Table 4. *The threshold evaluation using different features.*

| Feature | SP/CP (frame level) | SP/CP (segment level) | $\lambda_{act}$ |
|---|---|---|---|
| 12 MFCC + energy | 0.763/ 0.782 | 0.828/ 0.824 | 4.91 |
| 15 MFCC + energy | 0.765/ 0.787 | 0.831/ 0.805 | 4.42 |
| 12 MFCC + energy+ $\Delta$ MFCC | 0.750/ 0.739 | 0.840/ 0.783 | 2.28 |
| 12 MFCC + energy+ $\Delta\Delta$ MFCC | 0.768/ 0.710 | 0.860/ 0.765 | 1.88 |

## 8. Conclusions

The paper describes the method for fast speaker indexing. To obtain an effective and reliable clustering we use a two-level method. In the first stage of the merging process we apply a computationally cheap method based on the VQ which can effectively and correctly select audio segment candidates for merging and then in the second stage to use a more computational expensive technique based on the BIC for segment merging. The speed of the clustering is much quicker than the duration of the processed data. We also suggest the on-line procedure to provide automatically the data dependent value of the threshold for the BIC method. The experiments are carried out using 10 hours of the corpus collected in the French national project ESTER. The new system based on a two stage approach provides the results that are comparable with the results of the state-of-the-art clustering algorithm based in the BIC. The experiments demonstrate that the suggested system is 5 times (depending on the data) faster than baseline system and 20 times faster than real-time without affecting speaker indexing error rate. The suggested technique for the data dependent BIC threshold calculation using clustering data is effective for the data with the different features and feature dimensionality. The considered approach does practical speakers and audio clustering in a large audio database.